# Ionization of small molecules induced by H⁺, He⁺ and N⁺ projectiles: comparison of experiment with quantum and classical calculations

S. T. S. Kovács, P. Herczku, Z. Juhász, L. Sarkadi, L. Gulyás, B. Sulik

*Institute for Nuclear Research, Hungarian Academy of Sciences (ATOMKI), PO Box 51, H-4001 Debrecen, Hungary*

We report the energy and angular distribution of ejected electrons from $CH_4$ and $H_2O$ molecules impacted by 1 MeV H⁺, He⁺ and 650 keV N⁺ ions. Spectra were measured at different observation angles, from 2 eV to 2000 eV. The obtained absolute double-differential-electron-emission cross sections (DDCS) were compared with the results of CTMC and CDW-EIS calculations. For the bare H⁺ projectile both theories show remarkable agreement with the experiment at all observed angles and energies. The CTMC results are in similarly good agreement with the DDCS spectra obtained for impact by dressed He⁺ and N⁺ ions, where screening effects and electron loss from the projectile gain importance. The CDW-EIS calculations slightly overestimate the electron loss for 1 MeV He⁺ impact, and overestimate both the target and projectile ionization at low emitted electron energies for 650 keV N⁺ impact. The contribution of multiple electron scattering by the projectile and target centers (Fermi-shuttle) dominates the N⁺-impact spectra at higher electron energies, and it is well reproduced by the non-perturbative CTMC calculations. The contributions of different processes in medium velocity collisions of dressed ions with molecules are determined.

## INTRODUCTION

Ionization of atoms and molecules by ion impact has been extensively studied for decades. Cross sections for electron emission, differential in electron energy and emission angle are rich in structures, providing valuable information about the different ionization mechanisms. Covering the field of heavy particle – atom collisions, the main concepts, experimental methods and theoretical works for electron emission have been reviewed by Stolterfoht, DuBois and Rivarola in 1997 [1]. In the last two decades, remarkable new developments have been made in studying coherent electron emission from simple molecules by ion impact [2, 3], and collision systems with strong perturbation [4, 5], where multiple ionization [6-8] or multiple electron-scattering (Fermi shuttle) [9-13] significantly contribute. Ionization in ion-molecule collisions have also been studied from the beginning of electron spectroscopy [14]. Rather early, a remarkable systematic work has been performed by Wilson and Toburen [15]. They found that the ionization cross section of small carbohydrates by proton impact is scalable with the number of weekly bound electrons [15, 16].

Recently, the interest is rapidly increasing for studying ionization processes in molecular systems. For ion impact ionization, it is partially motivated by the rising applications, like plasma physics, ion therapy, and radiation protection as well as functionalization of polymers and other materials. Motivation for understanding the molecular-collision-governed large-scale processes in space and in planetary atmospheres (with a strong relevance of climate research) is also getting stronger. Last but not least, the dynamics of strongly perturbed molecules is a challenge for theories. Molecular ionization is often followed by the dissociation of the molecule. The connection between ionization and fragmentation is a subject of numerous studies and is intensively investigated [17-19].

In recent studies attention is turned to more complex systems, for instance collisions with large biomolecules [16, 20-22]. The experiments cover a wide energy range from few keV up to cca. 100 MeV. Ionization by bare ions was investigated in details by Tribedi et al. for a wide range of targets with emphasis on the two-center (TCE) and interference effects [21-25].

Collisional studies on biomolecules are especially interesting for hadron therapy [26], where large scale irradiation dose and cell damage model calculations need a big amount of reliable input data, including experimentally obtained ionization cross sections. Numerous measurements have been focused on the direct ionization of nucleobases [16, 20-22]. Although the direct DNA damage is of particular interest, secondary processes significantly contribute to the DNA lesion. Free electrons from primary ionization may induce molecular fragmentation by subsequent ionizations or by dissociative electron attachment (DEA) [27, 28]. Accordingly, differential electron production cross sections are important for mapping radiation damages on their own right.

In the majority of the applications based on ion-matter interactions, including radiation damages in biological tissues, the most important projectile energy region is where the linear energy transfer maximizes. This is known as the Bragg-peak region, centered typically at a few hundred keV/u impact energy. In this energy range, the equilibrium charge state of the ion inside the matter, which is determined by the relative yields of the charge transfer and ionization, is close to unity ($q \approx 1$) [29]. Studies in this energy region on small, few-atomic molecules with singly charged heavy ion projectiles are rather scarce. Montenegro et al. measured the ion production in C⁺+$H_2O$ collision from 15 keV to 100 keV impact energy [29]. However, according to our knowledge, double differential electron emission from molecules by singly charged, heavier ion projectiles in the Bragg-peak region has not been studied yet.

Electron spectra from collisions with dressed heavy ions exhibit the signatures of many ionization mechanisms, including the electron loss contributions. Single ionization is dominated by far collisions therefore it is governed by the screened ionic charge. In close collisions, however, screening effects are less important. Therefore multiple ionization cross sections are more close to those for bare projectiles [1, 30-32]. For dressed projectiles, multiple electron scattering at lower energies [9-13] and dielectronic (anti-screening) effects [33] at higher energies may also provide observable contribution to the spectra. Strong angular asymmetries due to two- or many-center effects are also expected in the studied energy region [21, 23, 25, 34]. Altogether, in the actual energy and projectile charge-state region of the present study, a full treatment of the whole ion-molecule collision scenario represents a challenge for collision theories.

In the present work, the predictions of two theoretical approaches are compared with experiment. On the quantum mechanical side, the continuum distorted wave, eikonal initial state (CDW-EIS) model has been extended for screened ion potentials and molecular targets [7, 35]. Moreover, we generalize the non-perturbative classical trajectory Monte-Carlo method (CTMC) for treating molecular targets by accounting for the ionization from the particular molecular orbitals [36].

In this work we focus on collision systems relevant for the Bragg-peak region. We report double differential cross sections for electron ejection from $H_2O$ and $CH_4$ molecules colliding with singly charged projectiles with different atomic numbers between 46 keV/u and 1000 keV/u impact energies. The results are compared with quantum mechanical CDW-EIS and classical CTMC calculations. Both targets are considered as tissue equivalent materials [26] since more than 60% of the human body is water, and the rest consists of carbon-based compounds. Therefore, we expect the results as important not only for understanding the dressed ion impact in the intermediate energy region, but also being relevant for hadron therapy.

In the following, we start with a description of the experimental system and the data evaluation methods. Then the details of the extension of the CTMC theory for dressed projectile impact on molecular targets are provided. For the CDW-EIS method a shorter desription is given. In the results and discussion part, we concentrate on the differences in the double differential electron emission cross section for the three projectiles. We analyze the potentialities of the theoretical approaches in accounting for the contributions of the different ionization mechanisms quantitatively. Finally, the role of those contributions in the studied collision systems is summarized.

## EXPERIMENT

We investigated the ionization of $H_2O$ and $CH_4$ molecules in a standard crossed beam arrangement. The scheme of the experimental setup is shown in Figure 1. Beams of $H^+$, $He^+$ and $N^+$ ions were supplied by the 5 MV Van de Graaff accelerator in Atomki, Debrecen, with kinetic energies of 1 MeV/u, 250 keV/u and 46 keV/u respectively. For keeping the charge state of the ions well defined, the beam was electrostatically deflected by 15° before being collimated and sent to the collision region. The cylindrical experimental chamber of 1000 mm diameter was equipped with two rotating rings. After the beam passed through the target volume, its current was measured in a two stage Faraday cup. For setting the position of the ion beam, a circular aperture ( A1) of 2 mm diameter, mounted on one of the rings, was turned to its aligned position. The beam was then sent through this aperture and the circular 3 mm hole at the bottom of the first Faraday stage. The target current was measured in the second Faraday cup. The distance between the two apertures was 800 mm inside the chamber. After maximizing the target current, with practically zero current in the first Faraday stage, a four-jawed slit - located upstream in the beamline at ~1200 mm distance from the entrance of the chamber - was closed to 2x2 mm without losing target current. Finally, the first 2 mm aperture (A1) was turned out from the beam direction. The setting was accepted when the target current (in the second Faraday cup) slightly increased by the removal of A1, and no significant current appeared in the first Faraday cup. Between the chamber and the four-jawed slit, one more, somewhat larger aperture was applied (A2) for avoiding secondary electron scattering into the experimental chamber. With a negligible current in the first Faraday stage, the beam position was determined by 2 distant elements, the four-jawed slit and the bottom hole of the first Faraday stage.

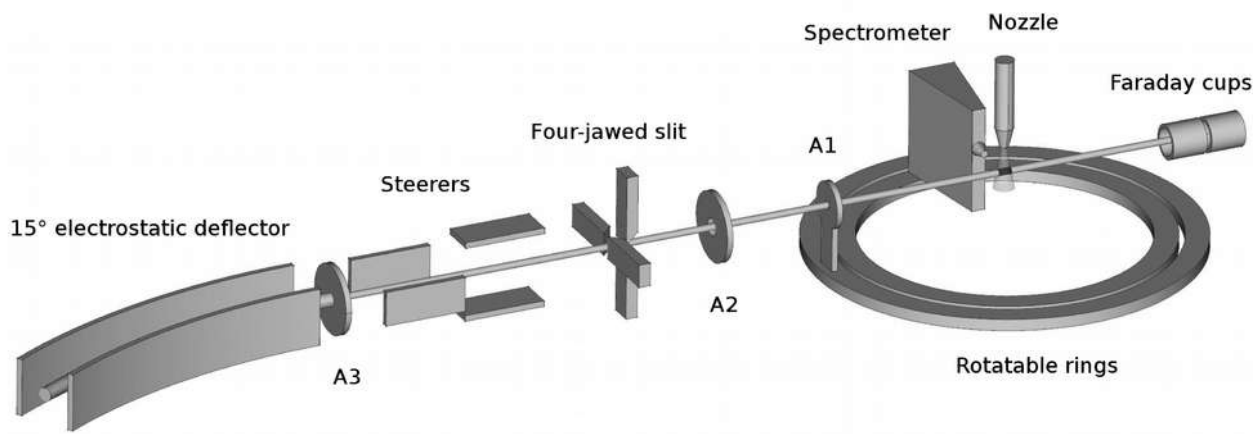

**Figure 1:** Schematic figure of the experimental setup.

With a double magnetic shielding the earth magnetic field was reduced below 2 mG everywhere in the relevant central part of the chamber. The residual gas pressure without gas target was lower than $10^{-6}$ mbar.

Jets of $H_2O$ and $CH_4$ gases were injected into the experimental chamber through a 1 mm diameter nozzle. A home-made actuator allowed us to move the nozzle away or towards the ion beam along the vertical axis of the cylindrical chamber. The gas flow was regulated by keeping a buffer pressure constant with the help of an automatically operated magnetic valve. The gas flow into the chamber was reduced by a long capillary. During the measurements the target gas density in the collision volume was estimated to be around $10^{13}$ cm$^{-3}$, which ensured single collision conditions. For water vapor target a liquid water reservoir was attached to the nozzle's pipe line. Before evaporating the pre-purified carbon free water into the vacuum chamber, dissolved gases were carefully pumped out from the liquid. The typical pressure in the scattering chamber was around $9 \times 10^{-7}$ mbar and $1 \times 10^{-5}$ mbar with and without gas inlet respectively.

Electrons, ejected from the crossing volume of the ion beam and the gas jet were energy analyzed by a single stage, parallel plate spectrometer. The special "axe" shape of it allowed us to move the spectrometer around the collision volume from 20° to 160° relative to the incident ion beam, so the measurements were performed for different observation angles within this range.

The distance from the collision volume to the channeltron detector was cca. 10 cm. The active volume from where the spectrometer collects the electrons is somewhat larger than the reaction volume (the crossing volume of the ion beam and gas beam), which provides a stability against small changes in the projectile and gas beams during the measurement. The base energy resolution of the spectrometer was 3%. Electron energy spectra were measured by scanning the analyzer voltage between 1.2 V and 1200 V in logarithmic increments. This corresponds to the energy region of 2-2000 eV. In each energy step the number of electrons transmitted by the analyzer was recorded for a fixed time period. The collected charge of the projectile beam in the Faraday cup was also measured in each energy steps.

For a reliable determination of the double differential cross section for electron emission, the effective target length ($L$) was calculated from the geometry of the collision volume, and the geometrical data of the spectrometer. It was approximated by a linear combination of two extremes as it is given in Ref. [37]:

$$L = L_0 [c + (1-c)\sin(\Theta)]^{-1} \qquad (1)$$

-----

Here $\Theta$ stands for the observation angle. The homogeneous target gas distribution along the projectile beamline is represented by $c = 0$, while $c = 1$ belongs to an ideal, dense, cylindrical target with small diameter. For our present measurements $c$ was determined by isotropically emitted ion spectra, and it was found to be $c = 0.6 \pm 0.03$.

Since it was not possible to measure directly the target density in the collision volume, for the normalization it was obtained by applying homogeneous gas pressure in the scattering chamber, similarly to the method of Ref. [37]. Without changing any other experimental parameter the nozzle was lifted up by 70 mm and the target gas flow was increased by a factor of 2-3 in order to keep the counting rate acceptable. The achieved homogeneous pressure was measured directly by an ionization gauge and was kept around $5 \times 10^{-5}$ mbar. Dividing the normalized electron emission yields for the known homogeneous pressure by those obtained in normal measurements we got a gain factor of G=51 at 90° observation angle. It means that the average pressure is about 51 times higher in the collision region below the nozzle in its normal position than the base pressure in the chamber during the measurement. Using this empirical factor from the continuously monitored chamber pressure we are able to estimate the average target gas density in the collision volume for normalization.

For the $H^+ + CH_4$ collision system at 1 MeV impact energy, absolute CS has been measured previously with good accuracy by Wilson and Toburen [15]. By assuming the detection efficiency of $\eta=0.7$ we successfully reproduced their reference results at all observation angles without any correction factors. In this way, we were able to perform absolute cross section measurements.

The reported accuracy of 20% for the reference data [15] represents a lower limit for the accuracy of our measured cross sections. For the electron spectra, the statistical uncertainty is increasing toward lower and higher energies, due to the decreasing count rate in both directions. In the 8-200 eV region it was 15% at most for $H^+$ impact. It was ~25% at 2 eV, and gradually increased above 200 eV, towards higher energies. For $He^+$ and $N^+$ projectiles, the statistical uncertainty was well below 10% at all angles in the 2-300 eV region.

Three sources of systematic errors were considered. Uncertainties in the determination of the effective target density and the collected charge are independent of the electron energy. Surface charging effects may influence the low energy part of the spectra. We estimate the systematic error around 40% at 2 eV, which decreases to 15% at 15 eV, and remains at that level elsewhere.

The overall uncertainty of the cross section data in the 5-300 eV energy range is ≤40%. At lower energies, it can approach 50-60%. At the high energy end of the spectra the statistical uncertainty dominates. Since in most regions the error bars are smaller than the size of the symbols, we do not show error bars in the figures, only at some high-energy points for demonstrating the increase of the statistical error, if relevant.

**THE CTMC MODEL**

Assuming the validity of the independent particle model (IPM), we applied a three-body CTMC approach that considers the interaction between the projectile, an active electron, and the ion core of the molecule. The CTMC method is based on the numerical solution of the classical equations of motion for a large number of trajectories of the interacting particles under randomly chosen initial conditions [38, 39].

The present CTMC computer code worked out for the description of the ion-molecules collisions is based on a previous code used for ion-atom collisions (for details see Ref. [13]). Our CTMC model in many aspects is very similar to that of Illescas *et al.* [40]. Unlike the latter authors, we describe the full three-body dynamics, i.e., we do not use the straight-line approximation for the projectile's path.

First we consider the water molecule. In the molecular orbital (MO) description of $H_2O$ the 10 electrons of the molecule occupy 5 MOs. The electrons in the lowest energy MO ($1a_1$) play a negligible role in the collision, so the active electron is chosen from the eight electrons in the four $2a_1$, $1b_2$, $3a_1$, and $1b_1$ valence MOs characterized by orbital energies -1.18, -0.67, -0.54, and -0.46 a.u. [41]. Each orbital contains two electrons. Assuming the validity of the Franck-Condon approximation, the calculations are carried out at fixed, equilibrium geometry of $H_2O$ defined by bond length of 1.811 a.u. and bond angle of 104.45°.

The active electron moves in the mean field created by the nuclei and the other nine electrons. Applying the three-center model potential proposed by Illescas *et al.* for the description of the mean field, the potential energy of the electron is expressed as

$$V_{\text{mod}}(\mathbf{r}) = V_O(r_O) + V_H(r_{H1}) + V_H(r_{H2}) \qquad (2)$$

with

$$V_O(r_O) = -\frac{8 - N_O}{r_O} - \frac{N_O}{r_O}(1 + \alpha_O r_O)\exp(-2\alpha_O r_O) \qquad (3)$$

$$V_H(r_H) = -\frac{8 - N_H}{r_H} - \frac{N_H}{r_H}(1 + \alpha_H r_H)\exp(-2\alpha_H r_H) \qquad (4)$$

where $r_O$, $r_{H1}$, and $r_{H2}$ in (2) are the electron distances to the three target nuclei, and the parameters in (3, 4) are: $N_O = 7.1$, $N_H = (9 - N_O)/2$, $\alpha_O = 1.500$ a.u., and $\alpha_H = 0.665$ a.u.

The trajectories of the particles are obtained by solving Newton's non-relativistic equations of motion:

$$m_i \frac{d^2 \mathbf{r}_i}{dt^2} = \sum_{j(\neq i)=1}^{3} \mathbf{F}_{ij}(\mathbf{r}_i - \mathbf{r}_j), \qquad (i=1,2,3). \qquad (5)$$

Here $m_i$ and $\mathbf{r}_i$ are the masses and the position vectors of the three particles, respectively. Introducing the notations e, P, and T for the electron, projectile, and target, the $\mathbf{F}_{ij}$ forces in (5) are the e-P, e-T, and P-T interactions. The e-T force is determined as $-\nabla_{\mathbf{r}_{ij}} V_{\text{mod}}(\mathbf{r}_{ij})$, where $\mathbf{r}_{ij} = \mathbf{r}_i - \mathbf{r}_j$ is the relative position vector of the two particles. For a bare ion projectile of charge $Z_P$ the P-T force is derived similarly: $-Z_P \nabla_{\mathbf{r}_{ij}}[-V_{\text{mod}}(\mathbf{r}_{ij})]$. The e-P interaction in this case is Coulombic. For a structured projectile ion we applied the Green-Sellin-Zachor (GSZ) potential [42] for the determination of the e-P and P-T interactions:

$$V^{\text{GSZ}}(r) = -\{Z - (N-1)[1 - \Omega(r, \eta, \xi)]\}/r \qquad (6)$$

where $Z$ is the nuclear charge, N is the number of the electrons in the ion, and

$$\Omega(r, \eta, \xi) = \{(\eta/\xi)[\exp(\xi r) - 1] + 1\}^{-1} \quad.$$

Here $\eta$ and $\xi$ are parameters determined by energy minimization. The e-P force is expressed as $-\nabla_{\mathbf{r}_{ij}} V^{\text{GSZ}}(\mathbf{r}_{ij})$. The P-T force was derived in the same way as for the bare ion projectile but with use of an effective ion charge $Z_P^{\text{eff}}$. The latter was obtained from the GSZ potential: $Z_P^{\text{eff}} = -r V^{\text{GSZ}}(r)$.

Our procedure for the generation of the initial values of the position and momentum coordinates of the electron from a set of uniformly distributed variables was different from that of Illescas et al. [40]. In our work we followed the method suggested by Reinhold and Falcón [43] for non-Coulombic systems. The latter authors considered the problem of central (isotropic) potential, we generalized their results for the non-isotropic potentials of the molecules.

Reinhold and Falcón started out from the basic assumption of the CTMC method [38], namely that the electron density in the phase space is constant,

$$f(\mathbf{r}, \mathbf{p}) = k\, \delta(E_i - E) \qquad (7)$$

where k is a normalization constant, $E_i$ is the ionization energy, and E is the total energy of the electron $E = p^2/2\mu + V(\mathbf{r})$ with $\mu = m_T/(1 + m_T)$. The $\delta$ function in ref. [6] ensures the energy conservation.

First we consider the case of the isotropic potential, i.e., $V(\mathbf{r}) = V(r)$. The range of $r$ is confined to the interval $0 < r < r_0$ because of the condition that the kinetic energy is positive. The maximum value $r_0$ is obtained as the root of the equation

$$E_i - V(r) = 0. \qquad (8)$$

We consider only potentials for which the above equation has one root.

Reinhold and Falcon arrived at a set of uniformly distributed variables by two successive variable transformations. The first one

$$(\mathbf{r}, \mathbf{p}) \rightarrow (E_i, r, v_r, v_p, \Phi_r, \Phi_p) \qquad (9)$$

is defined by

$$x = r(1 - v_r^2)^{1/2} \cos \Phi_r$$

$$y = r(1 - v_r^2)^{1/2} \sin \Phi_r$$

$$z = r v_r \qquad (10)$$

$$p_x = \{2\mu[E_i - V(r)]\}^{1/2}(1 - v_p^2)^{1/2} \cos \Phi_p$$

$$p_y = \{2\mu[E_i - V(r)]\}^{1/2}(1 - v_p^2)^{1/2} \sin \Phi_p$$

$$p_z = \{2\mu[E_i - V(r)]\}^{1/2} v_p \quad .$$

The second transformation is

$$(E_i, r, v_r, v_p, \Phi_r, \Phi_p) \to (\omega, v_r, v_p, \Phi_r, \Phi_p) \qquad (11)$$

Here the new variable $\omega$ is given by the integral

$$\omega(r) = \int_0^r dr' \mu r'^2 \{2\mu[E_i - V(r')]\}^{1/2} \qquad (12)$$

$\omega$, $v_r$, $v_p$, $\Phi_r$, and $\Phi_p$ are uniformly distributed variables in the intervals

$$\omega \in [0, \omega(r_0)], \quad \Phi_r \in [0, 2\pi], \quad \Phi_p \in [0, 2\pi],$$

$$v_r \in [-1, 1], \quad v_p \in [-1, 1].$$

Once a value of $\omega$ is chosen at random, $r$ is obtained from the inverse of the $\omega(r)$ function given by the integral (12), and the position and momentum coordinates are calculated using (9) with the randomly selected values of $v_r$, $v_p$, $\Phi_r$, and $\Phi_p$.

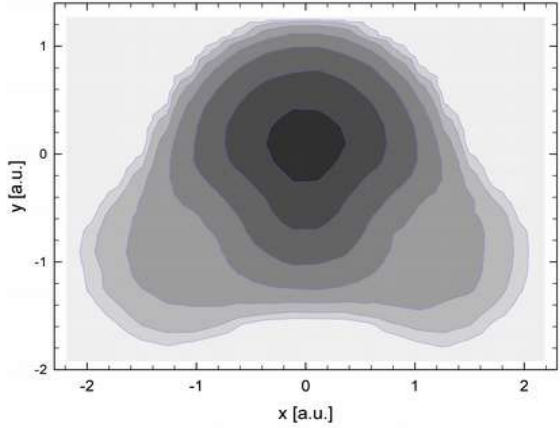

**Figure 2:** (Color online) Contour map of the probability density for of the initial electron position for the $2a_1$ MO of $H_2O$. Here $x$ and $y$ are coordinates of the projection of the position vectors into the molecule plane. From light to dark each intensity level increases by factor of two.

For non-isotropic potential $V(\mathbf{r})$ first we select $v_r$ and $\Phi_r$ that define the direction $(\theta_r, \Phi_r)$ of the $\mathbf{r}$ vector. (In Eqs. (10) one can identify $v_r$ as $\cos\theta_r$.) The difference from the case of the isotropic potential is that now the root of Eq. (8) has to be determined for each selected direction, i.e., $r_0 = r_0(\theta_r, \Phi_r)$. Also, the variable $\omega$ given by Eq. (12) has directional dependence:

$$\omega(r, \theta_r, \Phi_r) = \int_0^r dr' \mu r'^2 \{2\mu[E_i - V(r', \theta_r, \Phi_r)]\}^{1/2}$$
$$(13)$$

Again, after the random choice of an $\omega$ value with the condition $0 < \omega < \omega[r_0(\theta_r, \Phi_r)]$, r is obtained from the inverse of the $\omega(r, \theta_r, \Phi_r)$ function. Then the components of the momentum vector are calculated with the randomly selected values of $v_p$ and $\Phi_p$:

$$p_x = \{2\mu[E_i - V(r, \theta_r, \Phi_r)]\}^{1/2}(1-v_p^2)^{1/2}\cos\Phi_p$$

$$p_y = \{2\mu[E_i - V(r, \theta_r, \Phi_r)]\}^{1/2}(1-v_p^2)^{1/2}\sin\Phi_p$$

$$p_z = \{2\mu[E_i - V(r, \theta_r, \Phi_r)]\}^{1/2} v_p \quad . \qquad (14)$$

The distribution of the initial electron position calculated by the above-outlined procedure for the $2a_1$ MO of $H_2O$ is presented in Figure 2. The molecule lies in the $x$, $y$ plane, the contour map shows the projection of the position vectors into the molecule plane. In Figure 3 we compare the probability density function of the electron initial radial distance for the $2a_1$ MO calculated in the present work with that obtained by Illescas et al. [40].

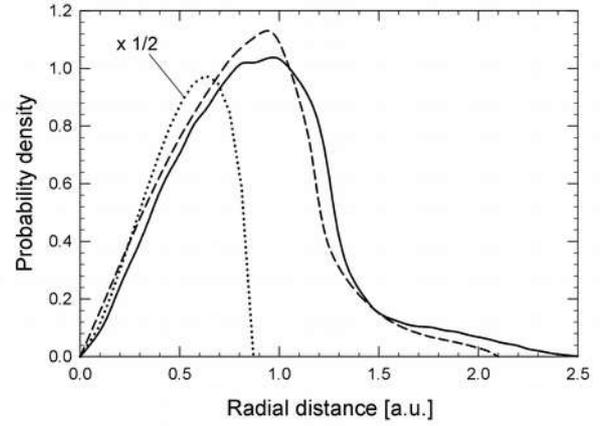

**Figure 3:** The probability density function of the electron initial radial distance for the $2a_1$ MO of $H_2O$. Solid line, present work; dashed line, Illescas et al. [40]; dotted line, the analytical result for $1/r$ potential and $E_i = 1.18$ a.u.

There is a reasonable agreement between the two distributions. Our procedure resulted in somewhat higher cut off value than that of Illescas *et al*. To see the difference between the atom and the molecule, we plotted also the analytical classical probability density function for an electron moving in a potential $V(r) = 1/r$ with the same ionization energy as that of the electron in the $2a_1$ MO.

For the molecule $CH_4$ a multi-center potential, similar to the three-center potential for $H_2O$, was not available in the literature. In our work we attempted to construct a five-center potential for $CH_4$. To do this, we analyzed the structure of the potential energy of the electron in $H_2O$ given by Eqs. (2)-(4). By substituting the numerical values of $N_O$ and $N_H$ into Eqs. (3)-(4), we obtain

$$V_O(r_O) = -\frac{0.9}{r_O} - \frac{7.1}{r_O} W(\alpha_O r_O), \qquad (15)$$

and

$$V_H(r_H) = -\frac{0.05}{r_H} - \frac{0.95}{r_H} W(\alpha_H r_H), \qquad (16)$$

with $W(x) = (1+x)\exp(-2x)$. Since the latter function decays exponentially, the second term in (15) and (16) can be identified as the short-range part of the potential energy.

The first term is the Coulombic, long-range part of the potential energy that determines the asymptotical behaviour of the (e + $H_2O^+$) system at large separation of the electron from the molecule core. For $r \to \infty$

$$V_{mod}(\mathbf{r}) = V_{mod}(r) = -\frac{0.9}{r} - 2\frac{0.05}{r} = -\frac{1}{r} \quad .(17)$$

At the same time, for the limit $r_O \to 0$ and $r_H \to 0$

$$V_O(r_O) = -\frac{8}{r_O}, \quad \text{and} \quad V_H(r_H) = -\frac{1}{r_H} \quad (18)$$

respectively. From Eqs. (15)-(18) we can conclude that the contribution of $V_O(r_O)$ to $V_{mod}(\mathbf{r})$ can be considered as that of an oxygen ion with ionic charge of almost one unit. Similarly, $V_H(r_H)$ is the potential energy of the electron interacting with an almost neutral hydrogen atom. This means that to a good approximation ($H_2O^+$ + e) ≈ ($O^+$ + e) + 2($H^0$ + e).

The above finding has led us to the idea that a multi-center potential can be constructed as a sum of atomic/ionic potentials with suitably chosen screening parameters. For a screened potential we may use the Green-Sellin-Zachor potential. $V^{GSZ}(r)$ can also be written as a sum of long- and short-range potential:

$$V^{GSZ}(r) = -\frac{Z-(N-1)}{r} - \frac{(N-1)}{r}\Omega(r,\eta,\xi) \quad . \quad (19)$$

Still remaining at $H_2O$, as a zeroth-order approximation we may apply the above potential for the $O^+$ ion and the $H^0$ atom, using $Z = 8$, $N = 8$ for $O^+$, and $Z = 1$, $N = 2$ for $H^0$. As a better approximation we may allow a small change of the electron number $N$ at each center: As is seen from Eq. (16), the hydrogen atoms are not completely screened, a small part of the electron cloud is moved from them to the oxygen ion. As a result, at the oxygen ion $N = 8 + \Delta N$, and at the hydrogen atoms $N = 2 - \Delta N/2$. In this approximation we have

$$V_O^{GSZ}(r_O) = -\frac{1-\Delta N}{r_O} - \frac{7+\Delta N}{r_O}\Omega(r_O,\eta_O,\xi_O) \quad (20)$$

$$V_H^{GSZ}(r_H) = -\frac{\Delta N/2}{r_H} - \frac{1-\Delta N/2}{r_H}\Omega(r_H,\eta_H,\xi_H) \quad . \quad (21)$$

The choice $\Delta N = 0.1$ in Eqs. (20) and (21) leads to the same long- and short-range potential coefficients as those in Eqs. (15) and (16).

For the generalization of the above result for other molecules one needs to know the value(s) of $\Delta N$. The latter quantity at a given atom in the molecule may be related to the partial charge defined as a difference between the charge calculated by a quantum chemical model around the atomic center and the charge of the neutral atom. The investigation of the correlation between the two quantities is beyond the scope of the present work, therefore for the construction of the five-center potential for $CH_4$ we assume that $\Delta N$ is small, and we can use the "zeroth-order approximation". As a check, we calculated the zeroth-order approximation for $H_2O$ using Eqs. (20) and (21) with $\Delta N = 0$. In Figure 4 we compared the approximate potential energy obtained in the direction of one of the hydrogen atoms with the result obtained by the three-center potential proposed by Illescas et al. [40].

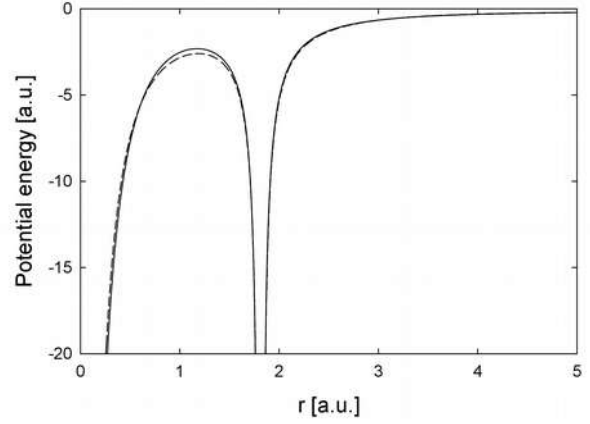

**Figure 4:** The potential energy of the electron in $H_2O$ in the direction of one of the hydrogen atoms. Solid line, the model potential given by Eqs. (2)-(4) [40]; dashed line, present result obtained in the zeroth-order approximation (see text).

The good agreement found between the two potential curves justifies the application of the zeroth-order approximation for $CH_4$. In this way in our CTMC calculations for $CH_4$ we used the following five-center potential:

$$V_{CH4}(\mathbf{r}) = V_C(r_C) + \sum_{i=1}^{4} V_H(r_{Hi}) \quad (22)$$

with

$$V_C(r_C) = -\frac{1}{r_C} - \frac{5}{r_C}\frac{1}{(\eta_C/\xi_C)[\exp(\xi_C r_C)-1]+1} \quad (23)$$

$$V_H(r_H) = -\frac{1}{r_H}\frac{1}{(\eta_H/\xi_H)[\exp(\xi_H r_H)-1]+1} \quad . \quad (24)$$

The parameters of the Green-Sellin-Zachor potentials in (23)-(24) were taken from Garvey et al. [44]. For carbon ($Z = N = 6$): $\eta_C = 2.13$ and $\xi_C = 1.065$; for hydrogen ($Z = 1$, $N = 2$): $\eta_H = 0.6298$ and $\xi_H = 1.3254$.

For $CH_4$ tetrahedral geometry with bond length of 2.067 a.u. was considered [45]. The electron emission was determined from the $2a_1$ and $1t_2$ valence MOs characterized with orbital energies -0.8452 a.u. and -0.5329 a.u. [46], as well as occupation numbers 2 and 6, respectively. The doubly differential cross section (DDCS) for the total electron production describing the energy and angular distribution of the electron following the ionization of the molecule is the sum of DDCSs for the electron emission from the individual MOs:

$$\frac{d^2\sigma}{d\varepsilon\, d\Omega} = \sum_{i=1}^{N_{MO}} \frac{d^2\sigma_i}{d\varepsilon\, d\Omega} \quad (25)$$

where $N_{MO}$ is the number of the MOs. For a given MO, classically the DDCS can be expressed as (omitting the subscript $i$)

$$\frac{d^2\sigma}{d\varepsilon\, d\Omega} = n\, 2\pi \int_0^\infty b\, \frac{d^2 p}{d\varepsilon\, d\Omega}(b)\, db \quad (26)$$

where $d^2p/d\varepsilon d\Omega$ is the one-electron doubly differential ionization probability for the regarded MO, $n$ is the number of the electrons in the MO, and $b$ is the impact parameter. For large number of collision events characterized by uniformly distributed b values in the range $(0, b_{max})$ the integral in (26) can be approximated by the following sum:

$$\int_0^\infty b\, \frac{d^2 p}{d\varepsilon\, d\Omega}(b)\, db \approx \frac{b_{max} \sum_j b_j}{N_{tot} \Delta\varepsilon\, \Delta\Omega}. \quad (27)$$

Here $b_j$ is the actual impact parameter at which the electron is emitted with energy and angle that falls in the energy window $\Delta\varepsilon$ and solid angle window $\Delta\Omega$, and $N_{tot}$ is the total number of the collision events.

For a fixed molecule orientation the electron emission depends on both the azimuthal and polar angle. In our work at each collision event the orientation of the molecule was randomly changed. This was achieved by the rotation of the molecule using the three Euler angles. By a suitable transformation of the Euler angles to uniformly distributed variables we ensured the isotropic distribution of the molecule orientation. Due to the resulting azimuthal symmetry of the electron emission, $\Delta\Omega$ is determined by the polar angular window ranging from $\theta_{min}$ to $\theta_{max}$:

$$\Delta\Omega = \int_0^{2\pi} \int_{\theta_{min}}^{\theta_{max}} \sin\theta\, d\theta\, d\Phi = 2\pi(\cos\theta_{min} - \cos\theta_{max}) \quad (28)$$

From Eqs. (27-28) we obtain for a given MO

$$\frac{d^2\sigma}{d\varepsilon\, d\Omega} \approx n\, \frac{b_{max} \sum_j b_j}{N_{tot}(\cos\theta_{min} - \cos\theta_{max})\Delta\varepsilon} \quad (29)$$

For a structured projectile the electron emission due to the ionization of the projectile has also to be taken into account. This can be done by considering the reversed collision system. For the electron production in collisions of nitrogen ions with water molecules the contribution of the projectile's ionization (known as "electron loss") was determined by calculating the process $H_2O + N^+ \rightarrow H_2O + N^{2+} + e$. We simplified the latter calculation by replacing the neutral $H_2O$ projectile by neutral oxygen atom. This can be justified by the dominant contribution of the oxygen to the molecular potential, as well as by the relatively small ionization probability of $N^+$ due to its large second ionization energy (-1.087 a.u.). In the same way for the $He^+ + CH_4$ collision the $CH_4$ projectile in the reversed system was replaced by neutral carbon atom. The results obtained for the reversed collision system refer to the projectile-centered frame for the direct system, therefore we had to transform them to the laboratory system. This was made simply by performing a velocity (Galilean) transformation on the trajectories.

**THE CDW-EIS MODEL**

Among the perturbative calculations the continuum distorted wave with eikonal initial states (CDW-EIS) approximation seems a powerful method to describe ionization of atoms by bare projectiles at medium and high impact energies [1, 23, 47]. In the present work an extended CDW-EIS method was applied to describe the electron emission from molecules impacted by dressed projectiles. The details together with the expressions can be found in Ref. [7, 35], so here we give only a draft of the method.

In the theoretical description we invoked the following three approximations: i) Impact parameter picture, where the incident projectile is assumed to move along a straight line trajectory $\mathbf{R} = \boldsymbol{\rho} + \mathbf{v}t$ with $\boldsymbol{\rho}$ perpendicular to $\mathbf{v}$, where the constant velocity $\mathbf{v}$ is parallel to the $z$ axis of the laboratory system fixed at the center of the molecule and $\boldsymbol{\rho} \equiv (\rho, \varphi_\rho)$ denotes the impact parameter. ii) Frank–Condon approximation, iii) IPM as defined for the description of CTMC.

The single-electron Hamiltonian is given by

$$h(\mathbf{x}, t) = h_0(\mathbf{x}) + V_p(\mathbf{s}) + V_s(\mathbf{R}(t)), \quad (30)$$

$$h_0(\mathbf{x}) = -\frac{1}{2}\Delta_x + V_{molecule}(\mathbf{x}) \quad (31)$$

where x denotes the position vector of the active electron with respect to the target center, $\mathbf{s} = \mathbf{x} - \mathbf{R}(t)$, $h_0$ denotes the electronic Hamiltonian for the target molecule, $V_{molecule}(\mathbf{x})$ describes the effective interaction of the electron with the target nucleus and other electrons, $V_p(\mathbf{s})$ is the interaction between the projectile and the active electron, and $V_s(\mathbf{R}(t))$ stands for the interaction of the projectile with the target nucleus and the passive electrons. Since $V_s(\mathbf{R}(t))$ depends only on the internuclear coordinate ($\mathbf{R}$), it can be accounted for by a phase factor. The latter does not affect the electron dynamics within the impact parameter approximation. Therefore, we drop this term in the following. In the same way as in the CTMC model, for $V_p$ here we also choose the GSZ potential [42] for the determination of the e-P and P-T interaction (cf. Eq. 19):

$$V_p(s) = V_p^s(s) - \frac{q}{s}, \quad \text{with} \quad q = Z - (N-1).$$

The transition amplitude for the ejection of electron with momentum $\mathbf{k}$ from the i-th initial orbital in prior form of CDW-EIS can be written as

$$a_{i\mathbf{k}}^-(\boldsymbol{\rho}, \omega_E) =$$

$$-i\int_{-\infty}^{\infty} dt \langle \xi_{\mathbf{k}}^{-}(\mathbf{x},t)|(\widetilde{h}(\mathbf{x},t)-i\frac{\partial}{\partial t})\xi_{i,\omega_E}^{+}(\mathbf{x},t)\rangle \quad , (32)$$

where $\omega_e$ denotes collectively the Euler angles (α, β, γ) referring to the orientation of the molecule. The wave functions $\xi^-$ and $\xi^+$ are given by eikonal distorted wave functions for the initial channel

$$\xi_{i,\omega_e}^{+}(\mathbf{x},t)=e^{-i\epsilon_i t}\Phi_{i,\omega_e}(\mathbf{x})E_{\mathbf{v}}^{*}(\mathbf{s},\eta_i) \quad (33)$$

and by Coulomb distorted wave functions

$$\xi_{\mathbf{k}}^{-}(\mathbf{x},t)=e^{-i\epsilon_k t}\Phi_{\mathbf{k}}^{-}(\mathbf{x})D_{\mathbf{p}}(\mathbf{s},\eta_p) \quad (34)$$

for the ionization channel. The distortion factors $E_{\mathbf{v}}(\mathbf{s},\eta_i)$ and $D_p(\mathbf{s},\eta_P)$ are given as

$$\begin{aligned}D_{\mathbf{p}}(\mathbf{s},\eta_p)&=e^{\pi\eta_p/2}\\&\times\Gamma(1+i\eta_p)_1F_1(-i\eta_p,1,-i(ps+\mathbf{p}\cdot\mathbf{s}))\end{aligned} \quad (35)$$

and

$$E_{\mathbf{v}}(\mathbf{s},\eta_i)=(vs+\mathbf{v}\cdot\mathbf{s})^{i\eta_i} \quad (36)$$

respectively, where $\eta_i = q/v$, $\eta_p = q/p$, **p=k-v**, **s=x-R**, and $_1F_1$ is the confluent hypergeometric function. $\widetilde{h}=h-V_s$ and

$$\begin{aligned}(\widetilde{h}-i\frac{\partial}{\partial t})&\xi_{i,\omega_E}^{+}(\mathbf{x},t)=-e^{-i\xi_i t}\\\times[\Phi_{i,\omega_E}&(\mathbf{x})\frac{1}{2}\nabla_x^2 E_{\mathbf{v}}^{*}(\mathbf{s},\eta_i)+\nabla_x\Phi_{i,\omega_E}(\mathbf{x})\cdot\nabla_s E_{\mathbf{v}}^{*}(\mathbf{s},\eta_i)\\+V_p^s&(s)\xi_{i,\omega_E}^{+}(\mathbf{x})]\end{aligned}$$
(37)

where the first two terms in the square brackets correspond to the transition due to the bare projectile impact, while the last term is due to the short range part of the projectile potential.

The initial atomic orbitals $\Phi_{i,\omega_E}$ are evaluated on the basis of the Hatree-Fock approach and have been described by using the STO-3G Gaussian basis set [50]. The expansion coefficients are obtained by the Gaussian computational chemistry software package [51].

As for the CTMC description, from the $(1a_1)^2$ $(2a_1)^2$ $(1t_2)^6$ electronic configuration of the $CH_4$ molecule only the $2a_1$ and $2t_1$ orbitals were considered, which had been constructed as linear combination of atomic orbitals. Similarly the deeply bound $(1a_1)^2$ orbital was excluded from the calculation in the case of $H_2O$, where only the $(2a_1)^2$ $(1b_2)^2$ $(3a_1)^2$ $(1b_1)^2$ orbitals were considered.

The continuum states of the molecule are described on a spherically averaged potential created by the nuclei and the passive electrons [52],

$$V_{molecule}^{+}(x)=V_{nuclei}(x)+V_{electrons}(x) \quad (38)$$

where

$$V_{electrons}(x)=-\sum_j^{N-1}\frac{1}{4\pi}\int d\mathbf{x}_1\frac{|\Phi_i(\mathbf{x}_1)|^2}{x'} \quad (39)$$

Here $x'=max(x_1,x)$, N is the number of electrons, and $V_{nuclei}(x)$ is represented by an averaged uniform spherical charge distribution for the nucleus according to their equilibrium distance from the center of molecule. As for the initial orbital, $\Phi_{\mathbf{k}}(\mathbf{x})$ is expanded over spherical harmonics,

$$\Phi_{\mathbf{k}}(\mathbf{x})=\frac{1}{x\sqrt{k}}\sum_{l,m}i^l e^{-i\delta_l}u_{k1}(x)[Y_1^m(\hat{\mathbf{x}})]Y_1^m(\hat{\mathbf{k}}) \quad ,(40)$$

where $u_{kl}(x)$ is obtained by the numerical solution for the radial part of the molecular Hamiltonian: $h_{molecule} = -\frac{1}{2}\Delta_x +V_{molecule}^{+}(x)$, see [53]. The probability for the electron emission with momentum **k** from a given $i$-th initial orbital is given by

$$p_{i,k_i}(\boldsymbol{\rho})=\frac{1}{8\pi^2}\int d\omega_E|a_{ik}^{-}(\boldsymbol{\rho},\omega_E)|^2 \quad , \quad (41)$$

where the integral over the Euler angles ($\int d\omega_E$) reflects to the conditions that the molecules have an arbitrary orientation in the experiments discussed in the next sections. The total ionization probability of the $i$-th initial orbital is obtained by integrating over **k**

$$p_i(\boldsymbol{\rho})=\int d\mathbf{k}\, p_{i,\mathbf{k}}(\boldsymbol{\rho}) \quad . \quad (42)$$

Up to this point we have discussed electron emission from the target, however, the projectile can also loose electron(s). As for the CTMC treatment, this process is treated similar to the target ionization if we change the reference frame from the target to the projectile one. In the projectile frame a neutral molecular particle as projectile with velocity **-v** ionize the ionic target. The potential for the molecular projectile is derived by the same way as for the atomic one but with the ground state configuration and orbitals of the negative molecule ion. The wave functions for the ionic target are evaluated by solving numerically the Schrödinger equation on the GSZ potential. Finally the derived transition probabilities are transformed back to the laboratory frame.

Having the single particle transition probabilities for the target and projectile, the probabilities for *q*-fold ionization are calculated by a binomial analysis within the framework of the independent particle picture. On the level of a shell-specific model [54] the probability of *q*-fold electron emission $p_{\mathbf{k}}^q$ when only one electron is ejected with momentum **k** is given by

$$p_{\mathbf{k}}^{(q)}(\boldsymbol{\rho})=$$

$$\frac{1}{m}\sum_{j:(1,2..m)}\sum_{\substack{q_1...q_m=0;\\q_1+...+q_m=q-1}}^{N_1,...N_m}\prod_{i=1}^{m}\frac{N_i!}{q_i!(N_i-q_i)!}p_{j,\mathbf{k}}(\boldsymbol{\rho})[p_j(\boldsymbol{\rho})]^{q_i}$$

$$\times[1-p_i(\boldsymbol{\rho})]^{N_i-q_i}$$
(43)

Here, *m* is the total number of electron shells on the projectile and target that can be ionized, and $N_i$ the number of electrons in each shell. Finally the doubly differential cross section for detecting one electron with energy $\varepsilon_k = \frac{1}{2}k^2$ in the direction of $d\Omega_k(\theta_k, \varphi_k)$ when $q$ electrons are removed from the projectile - target system is given by

$$\frac{d^2 p^{(q)}}{d\varepsilon_k d\Omega_k} = k \int d\boldsymbol{\rho}\, p_{\mathbf{k}}^{(q)}(\boldsymbol{\rho}) \qquad (44)$$

## RESULTS AND DISCUSSIONS

The measured double differential electron emission cross sections (DDCS) are displayed in figure 5 for the $CH_4$ target molecule. The figure exhibits the characteristic features of electron emission due to ion impact ionization of atoms or molecules. The cross section maximizes at low electron energies, where electrons originate from soft collisions. Between 10 and 1000 eV the cross section is decreasing by at least five orders of magnitude. Sharp target (carbon) and projectile (nitrogen) Auger peaks are visible in the spectra. The classical binary encounter peaks at $E_{binary}=4E_p \cos^2\theta$ energy, also appear at forward angles. They are best observable in the case of proton impact (Fig. 5 a). For the dressed $He^+$ and $N^+$ projectile ions, electron loss from the projectile (EL) also contribute to the spectra. The DDCS spectra of $H_2O$ are similar.

Among the characteristic features, the DDCS spectra exhibit significant differences for the three types of projectiles. Although their ionic charge is the same, the measured DDCS for $N^+$ is at least one order of magnitude higher than that for the $H^+$ projectile at all observed angles and electron energies. DDCS for $He^+$ lies between the above two. These differences might partially be attributed to the different impact velocities. The velocity ratios are 6.3 : 1.6 : 1.3 for $H^+$, $He^+$ and $N^+$ respectively. Indeed, if we characterize the interaction strength with the Bohr-Sommerfeld parameter as $\delta=q/v$, ($q$ is the ionic charge and $v$ is the velocity of the projectile) the strongest perturbation belongs to the $N^+$ projectile ($\delta_H=0{,}16$ $\delta_{He}=0{,}63$ and $\delta_N=0{,}74$). Moreover, the different nuclear charges also play a role in close collision events, where the projectile electrons may not screen the nuclear charge completely, and therefore the effective charge may highly exceed the ionic value for a short time period. Close encounters belong to large momentum and energy transfer. Accordingly, one may expect that the effective charge increases with the energy of the ejected electron. In close collisions the maximum value of $\delta$ remains the same for the bare $H^+$ ion, but it can be even doubled for $He^+$ ($\delta_{max}=1.26$). In the case of $N^+$ the maximum effective charge and the maximum value of the corresponding Bohr-Sommerfeld parameter can be as high as $q_{max}=7$ and $\delta_{max}=5.2$, respectively. We will see later that the increase of the cross section with electron energy can go even beyond a $Z^2$ scaling. Close collision events also cause multiple ionization of the target with large probability. We note that in Fig. 5b) and 5c), the EL process, i.e., the emission of electrons originating from projectile ionization remarkably contributes to the spectra.

For the 1 MeV $H^+$ projectile the angular distribution of low energy electron emission is close to isotropic. In contrary, strong forward-backward asymmetry in low energy electron emission has been found for $He^+$ and $N^+$ projectiles. This asymmetry originates from two sources. One is the enhanced two-center effect (TCE). Since the ionic charge is the same for all projectiles, the stronger asymmetry is partially attributed to the lower velocity of the heavier projectiles. Another source of the asymmetry is that in the laboratory frame, electrons ejected from the projectile are emitted dominantly into forward angles, enhancing the yield of low energy electrons, especially strongly for the $N^+$ projectile where the EL peak is centered at low energy (~25 eV).

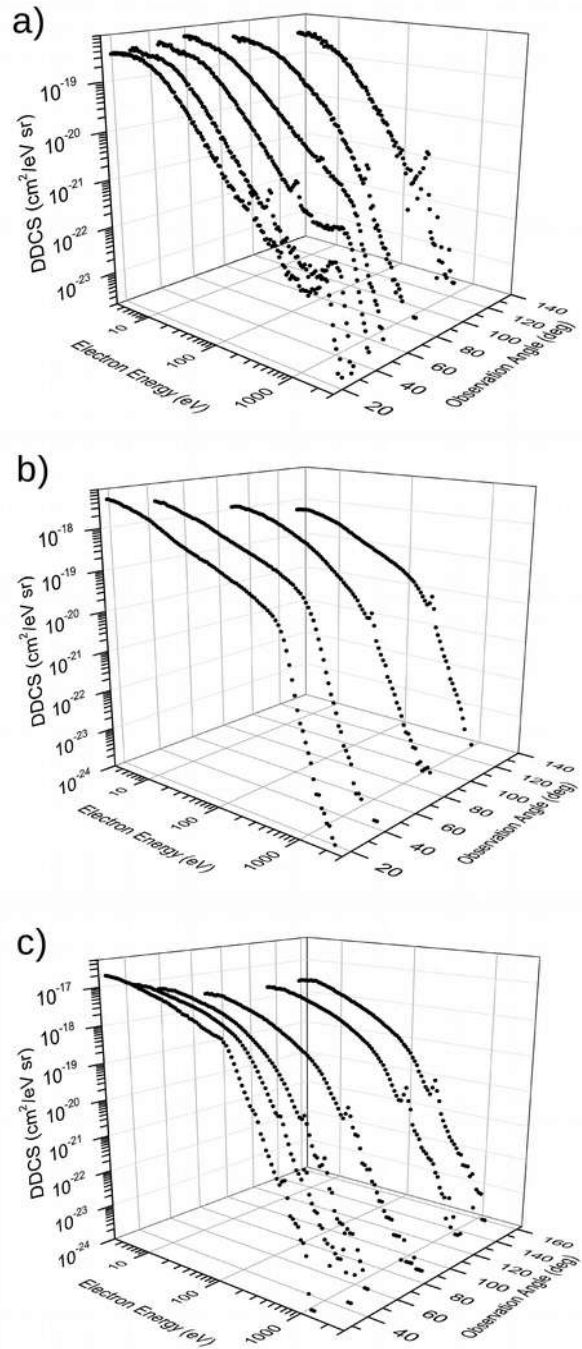

**Figure 5:** Double differential electron emission cross section (DDCS) for $CH_4$ target molecules induced by a) 1 MeV $H^+$; b) 1 MeV $He^+$; c) 650 keV $N^+$ projectiles. Spectra are depicted as a function of electron energy for different observation angles.

The measured DDCS spectra for the collision systems of $H^+ + CH_4$ and $H^+ + H_2O$ show remarkable agreement with both the CTMC and CDW-EIS calculations as it is shown in Figure 6 and 7. The shape of the spectra including that of the binary encounter peak is well reproduced. A small systematic difference between experiments and theories appear only below 10 eV where the experimental cross sections are slightly smaller than the theoretical predictions. Since our measured 5 eV data are also somewhat smaller than those of ref. [15], while the agreement is perfect above 10 eV, the deviation from the theories may partially be attributed to a systematic error in our measurement. We note, however, that the difference between the theories for $CH_4$ target below 10 eV and above 45° is larger than that between the CTMC results and our experimental data. In general, the agreement is very good between 10 and 100 eV. At larger electron energies the continuum distorted wave method slightly underestimates the experimental DDCS at 20° and 30° observation angles for $CH_4$. For both targets CTMC gives almost perfect agreement.

For the partially dressed $He^+$ and $N^+$ projectiles the theories have to take into account the electron loss contribution originating from the ionized projectile (EL). Figure 8a) and 8b) display the experimental data for $He^+ + CH_4$ collision compared with the calculated values by the CTMC and CDW-EIS theories, separately. Here we show not only the contribution from the ionization of the target, but also the total theoretical electron emission cross section, which contains both the target ionization and the EL contribution. Dielectronic interactions between the electrons of the projectile and the target [1] were estimated to be negligible here. Though the target ionization is dominant for $He^+$ impact at most observation angles, the results of the CTMC calculations in figure 8a) show a remarkable contribution of projectile ionization at backward angles. For $He^+$ impact, the target ionization is dominant due to the larger number of more loosely bound target electrons of $CH_4$ ($IE_I \approx 12.6$ eV) contrary to the more tightly bound single electron of $He^+$ ($IE_{II} \approx 54.4$ eV).

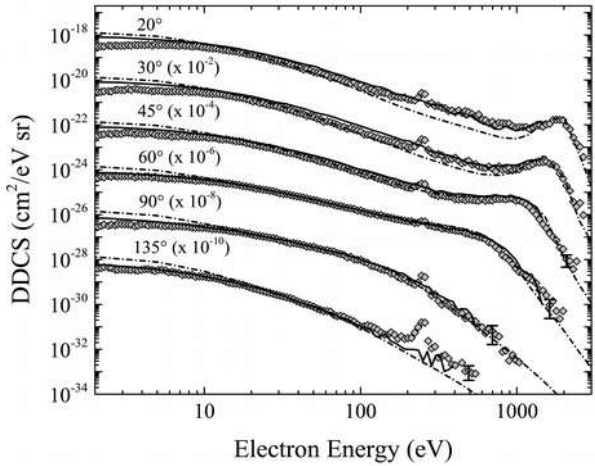

**Figure 6:** Double differential electron emission cross section for $CH_4$ by 1 MeV $H^+$ impact at different observation angles. The multiplying factors, applied for graphical reasons, are shown in parenthesis. The measured data are indicated as gray diamonds. The calculated CDW-EIS and CTMC values are depicted by dash-dotted and solid lines, respectively. A few sample error bars are shown at higher electron energies. At lower energies, the size of the diamonds exceeds the uncertainties of the data.

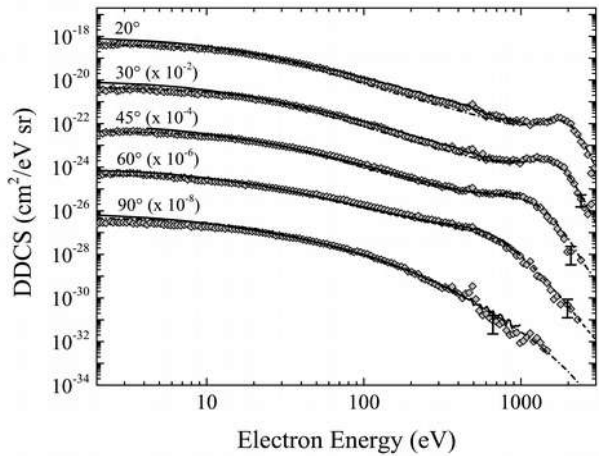

**Figure 7:** Double differential electron emission cross section for $H_2O$ by 1 MeV $H^+$ impact compared with CDW-EIS and CTMC calculations. The notations and comments are the same as for Fig 6.

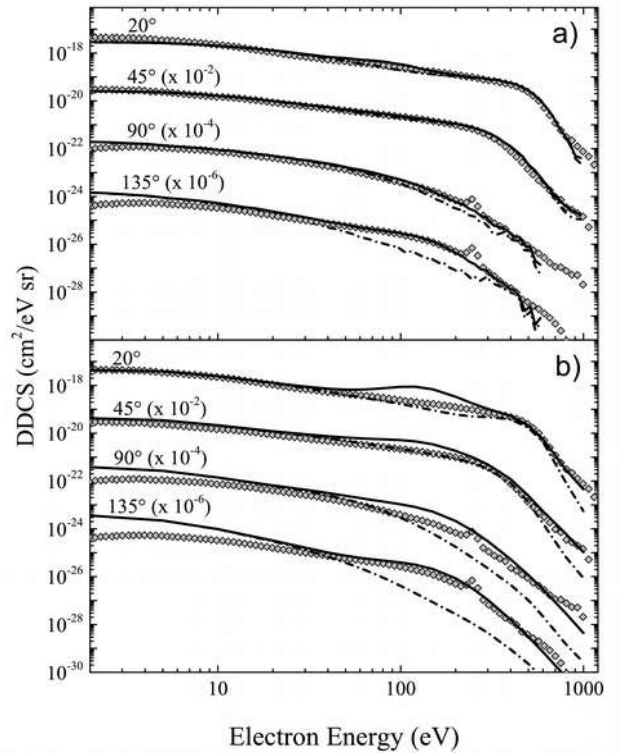

**Figure 8:** Double differential electron emission cross section for $CH_4$ in collisions with 1 MeV $He^+$ at different observation angles. The measured data are indicated as gray diamonds. The calculated target and total (target + projectile) ionization values for CTMC are depicted in part a) by dash-dotted and solid lines, respectively. In part b) the CDW-EIS values are presented with the same notation. The multiplying factors are shown in parenthesis.

Figure 8 b) shows that CDW-EIS overestimates, in some extent, the yield of projectile ionization at all observation angles. This is probably due to a slight overestimation of the contribution of screened potentials as ionizing 'agents'. Although small deviations remain mostly below 10 eV, the total (target + projectile) ionization values obtained by CTMC are in an almost perfect agreement with the experiment. It is noted that both theoretical models predict too large yield of low energy electrons emitted to the backward direction. This is a signature of underestimating the role of two-center effects.

The experimental results together with the calculated CTMC and CDW-EIS data for the collision system of $N^+$ +$H_2O$ are shown in Figure 9 a) and 9 b), respectively. Since the $N^+$ projectile carries much more electrons than $He^+$, and its ionization potential is about half of that of $He^+$, significantly higher projectile ionization yields can be expected. Indeed, according to the measurements, the absolute cross section for projectile ionization is more than 7 times higher for $N^+$ than that for $He^+$.

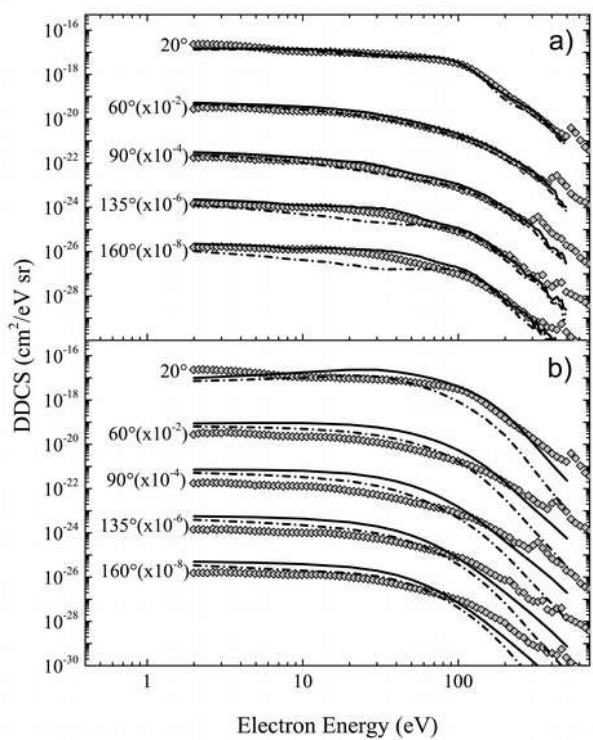

**Figure 9:** Double differential electron emission cross section for $H_2O$ in collisions with 650 keV $N^+$ at different observation angles. The multiplying factors are shown in parenthesis. The measured data are indicated as gray diamonds. In part a) CTMC results for target and total (target + projectile) ionization are displayed by dash-dotted and solid lines, respectively. In part b) experiment is compared with CDW-EIS results for target and total ionization with the same notations.

Differences between the non-perturbative classical and the distorted wave quantum mechanical results are more significant for the many-electron $N^+$ projectiles (see Figure 9). The CTMC calculation shows similar behavior as that for the $He^+$+ $CH_4$ collision system: at forward angles the contribution of the target ionization is dominant. With increasing observation angles the electron emission from the projectile gets importance. While the agreement between the experiment and the total ionization obtained by CTMC remains very good for the $N^+$ + $H_2O$ collision system, the CDW-EIS data overestimate the electron emission from both collision partners at low energies. One should note here, however, that at this impact energy and perturbation strength, the level of agreement is excellent for a basically first order perturbative treatment, demonstrating the power of the sophisticated handling of the initial and final states. The deviation of the CDW-EIS prediction from the experiment is more stringent at higher emitted electron energies, where the theoretical curves fall off much faster with electron energy than the experiment.

The experimental data for the 650 keV $N^+$ + $CH_4$ collision system together with the corresponding CDW-EIS results are displayed in Figure 10. They are similar to those of the 650 keV $N^+$+ $H_2O$ system. A closer inspection shows that the measured cross sections are almost equal in magnitude, and exhibit similar angular and energy dependence. Let us recall the scaling rule of Wilson and Toburen for proton impact on hydrocarbon molecules [15], namely that the electron ejection cross section is roughly proportional to the loosely bound electrons in the target molecule. Our present measurements suggest that this approximate scaling rule has a wider region of validity: It works for other molecules and for projectile ionization too.

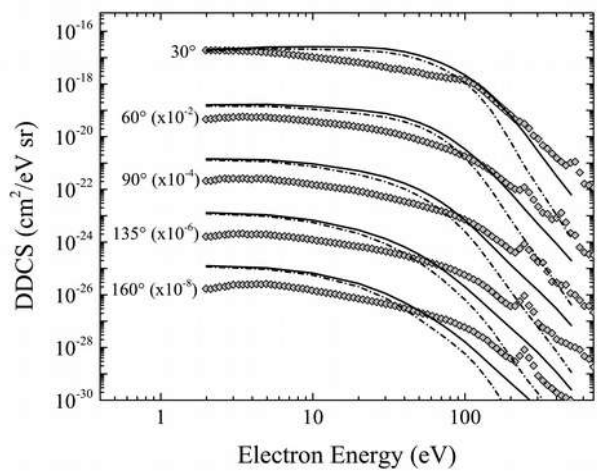

**Figure 10:** Double differential electron emission cross section for the 650 keV $N^+$ + $CH_4$ collision at different observation angles. The multiplying factors are shown in parenthesis. The measured data are indicated as gray diamonds and are compared with the CDW-EIS results. The calculations for the target and the total (target+projectile) ionization are displayed by dash-dotted and solid lines, respectively.

The experimental results for $N^+$ impact exhibits much larger cross sections above 100 eV electron energy than those predicted by the CDW-EIS model. A closer inspection shows a hump in the spectra around 100 eV, even at backward angles (see Figures 9 and 10). For the water target, CTMC reproduces well the high energy behavior of the cross section, moreover, it also predicts a hump at 100 eV in the target ionization contribution at backward angles. This is a key for understanding the origin of the hump. We assume that both the structure at 100 eV and the enhancement at higher energies in general is due to sequential multiple electron scattering in the Coulomb field of the ionized target and the projectile cores, i.e. the so called Fermi-shuttle mechanism [9-13].

A possible scenario for this mechanism in ion-atom collisions starts with a close encounter between a target electron and the incoming projectile core (P) in the approaching phase of the collision. If the electron is scattered forward in the laboratory frame, it has a chance to get scattered back in a close collision with the target core (T), and be scattered back by the projectile again (P) before it gets liberated. This scattering series can be characterized by the sequence of the cores where the electron gets scattered, e.g. for the above process by P-T-P. Similar scattering sequences can start with the ionization of the projectile, e.g., a T-P-T-P sequence. It is important that, due to the large projectile/electron mass ratio, the velocity of the electron increases by two times the velocity of the incoming projectile, 2V, in every P scattering. Four-fold scattering events have been first found in 1.8 MeV $C^+$ + Xe collisions in Ref. [9].

It has been found earlier [9-13] that, due to its non-perturbative character, the CTMC model was able to quantitatively treat this type of multiple scattering events in ~100 keV/u collisions. From the CDW-EIS model, however, which includes only parts of higher order perturbation, we do not expect to describe this type of multiple scattering. Accordingly, the dramatic underestimation of the cross section for $N^+$ impact by the CDW–EIS theory at higher electron energies on one side, and the good agreement between experiment and CTMC for the same collision system everywhere (see Figures 9 and 10) allows us to conclude that Fermi-shuttle type multiple scattering could significantly contribute to the high energy electron emission there.

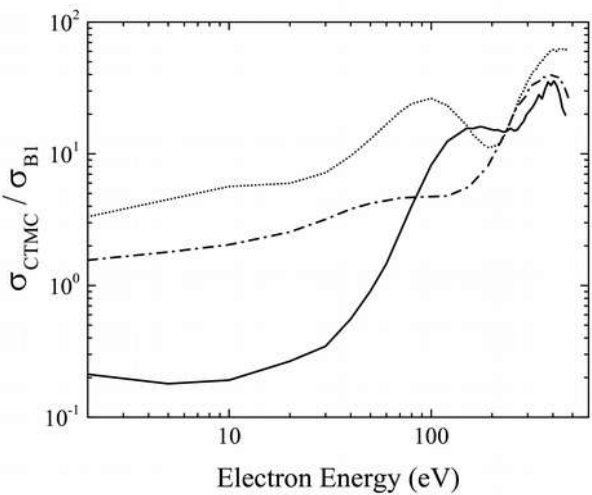

**Figure 11:** The CTMC calculations for 46 keV/u $N^+$+$H_2O$ collision divided by the result of B1 approximation for 46 keV/u $H^+$+$H_2O$ collision system at different observation angles: 20° - dotted line; 45° - dash-totted line; 160° - solid line. The ratios highlight the contribution of higher order processes as wide humps at higher energies. Multiple electron scattering (Fermi-shuttle) gives rise even at backward angles.

This conclusion is confirmed by the analysis given in Figure 11. Here the target ionization contribution calculated by the CTMC theory for the 46 keV/u $N^+$+$H_2O$ collision system is divided by the reference cross section of a first order plane-wave Born calculation for the equi-velocity 46 keV/u $H^+$+$H_2O$ collision system. The latter represents a pure first-order description of ionization by a bare unit charge at the same velocity without two-center contribution or higher order processes. In a simple picture, the ratio should be unity at low electron energies, where the ionic charge of the projectile should govern the cross section in distant collisions, and it should gradually increase towards higher energies, since the screening of the projectile nucleus decreases with closer and closer collisions. The ratio in Fig. 11 clearly shows that two-center effects dominate electron emission at energies below 30 eV. At forward angles it far exceeds unity, while it gets very small for backward emission. At higher energies, the structures nicely demonstrate the features of close collisions. The 20° curve exhibits a wide peak at 90 eV, corresponding to the enhancement of the binary encounter process due to the screened potential of the projectile [55]. Note that the zero-degree binary encounter energy is 25 eV here. This effect is smaller but still observable at 45° and ~50 eV. The main contribution to the peak above 100 eV in the 160° ratio is a P-T-P scattering sequence. Finally there is a strong peak-like enhancement around 400 eV for all observation angles. They can be identified as P-T-P and P-T-P-T (at 160°) processes, i.e., triple and quadruple scattering sequences on the nitrogen and oxygen cores (scattering on the protons in $H_2O$ should be negligible). Since experiment and CTMC agree well in this energy region, the ratios in Figure 11 provide an indirect evidence for the importance of multiple electron-scattering in our slowest (46 keV/u) collision system.

The shape of the EL peak is rather different in the two theoretical models. In both calculations, the electron loss from the $He^+$ projectile is peaking between 100 and 300 eV and gives only a slight contribution to the cross section spectra at other energies. However, EL obtained by CDW-EIS indicates a larger contribution and broader energy distribution of such electrons than those obtained by CTMC, even for $He^+$ impact. For $N^+$ projectile this peak turns into a broad flat hump lying between 10 and 70 eV in the measured spectra. Again, it is well reproduced by CTMC. Instead of a broad hump the CDW-EIS method gives an almost continuous EL contribution to the whole spectra. The limits of the extended CDW-EIS calculation here might be attributed to the treatment of the screened Coulomb potential of the $N^+$ ion.

## CONCLUSIONS

We carried out measurements for the ionization of $H_2O$ and $CH_4$ molecules by the impact of MeV energy $H^+$, $He^+$ and $N^+$ projectiles. The energy and angular distribution of the ejected electrons were measured by a single stage electrostatic spectrometer. The obtained absolute double differential electron emission spectra were compared with those of calculated by CTMC and CDW-EIS models, specially developed for treating screened ionic core potentials and accounting for the molecular orbitals of the target. We found that the measured cross sections increased with the increase of the atomic number Z of the projectile and the decrease of the projectile velocity. The spectral features in the case of the partially dressed projectiles could be attributed to the increased effective charge in close collision events, and the contribution from the ionization of the projectile.

A good general agreement was observed between measurement and extended CTMC calculations for all the studied collision systems, even for $N^+$ impact. Here multiple electron scattering have been found to dominate the high-energy part of the spectra. The CTMC model has been found a reliable tool for treating molecular collisions in all collision systems studied here at and below the mean energy of the Bragg peak.

The advanced CDW-EIS model provided good agreement with proton impact data, a reasonable agreement for $He^+$ impact, and a qualitative description of the low energy part of the $N^+$ impact data. Since CDW-EIS is basically a first order perturbation theory, this level of agreement clearly demonstrates the effectivity of applying realistic initial and final states. Since it was found to be less efficient for describing the ionization of the projectile by the completely screened target, it is likely that it should be developed further for treating the screened potentials.

The shape and the absolute value of the corresponding electron emission spectra are similar for $H_2O$ and $CH_4$. Since both molecules has 8 weekly bound electrons, the similarity suggest that the approximate rule of Ref [15] has a wide region of validity. The present measurements suggest that the ionization cross section is scalable with the number of loosely bound electrons in a target molecule, and this rule seems to be valid for different projectile ions at different impact energies.

Although, according to both theories, single ionization is dominant for the studied collision systems, at $N^+$ impact they predict strong contributions from double and multiple ionization, too. This can be verified by measuring the relative yields of the fragmentation channels of the target molecules in the same collision systems.

## ACKNOWLEDGMENTS

This work has been supported by the Hungarian Scientific Research Foundation (OTKA Grant No.: K109440), the National Information Infrastructure Program (NIIF) and by the TÁMOP-4.2.2B-15/1KONV-2015-0001 project. The authors thank the VdG-5 accelerator staff for the operation of the machine during the experiments.